\documentclass[conference]{IEEEtran}
\usepackage{graphicx, times, amsmath, amsfonts, amsthm, comment}
\usepackage{amssymb,epstopdf}
\usepackage{algorithmic}
\usepackage{algorithm}
\usepackage{color, soul}
\usepackage{mathtools}
\usepackage{caption}
\usepackage{subcaption}
\DeclarePairedDelimiter{\ceil}{\lceil}{\rceil}

\newcommand{\beq}{\begin{equation}}
\newcommand{\eeq}{\end{equation}}
\newcommand{\tbf}{\textbf}
\newcommand{\tit}{\textit}

\newcommand*{\mathcolor}{}
\def\mathcolor#1#{\mathcoloraux{#1}}
\newcommand*{\mathcoloraux}[3]{%
  \protect\leavevmode
  \begingroup
    \color#1{#2}#3%
  \endgroup
}

\newcommand {\Ebb}{\mathbb{E}}

\newcommand {\Acal}{\mathcal{A}}
\newcommand {\Fcal}{\mathcal{F}}

\newcommand {\Kcal}{\mathcal{K}}
\newcommand {\Lcal}{\mathcal{L}}

\newcommand {\Pcal}{\mathcal{P}}
\newcommand {\Scal}{\mathcal{S}}

\begin{document}

\title{Network Slicing with Mobile Edge Computing for Micro-Operator Networks in Beyond 5G}

\author{
\IEEEauthorblockN{Tachporn Sanguanpuak\IEEEauthorrefmark{1}, Nandana Rajatheva\IEEEauthorrefmark{1}, Dusit Niyato\IEEEauthorrefmark{2},  Matti Latva-aho\IEEEauthorrefmark{1}}
\IEEEauthorblockA{\IEEEauthorrefmark{1}Centre for Wireless Communications (CWC), University of Oulu, Finland; \\ \IEEEauthorrefmark{2}School of Comp. Science and Eng., Nanyang Technological University (NTU), Singapore}
\IEEEauthorblockA{Email: \{tachporn.sanguanpuak, nandana.rajatheva, matti.latva-aho\}@oulu.fi; dniyato@ntu.edu.sg}
}\maketitle

\begin{abstract}
We model the scenarios of network slicing allocation for the micro-operator (MO) network. The MO creates the slices ``as a service'' of wireless resource and then allocates these slices to multiple mobile network operators (MNOs). We propose the slice allocation problem of multiple MNOs with the goal of maximizing the social welfare of the network defined as sum rate of all MNOs. The many-to-one matching game framework is adopted to solve this problem. Then, the generic Markov Chain Monte Carlo (MCMC) method is introduced for the computation of game theoretical solution. After the MNOs obtain the slices, for each small cell base station (SBS), we investigate the role of power allocation using Q-learning and uniform power. We numerically show that the solution of the matching game leads to two-sided stable matching. Furthermore, for each MNO, we explore the problem of infrastructure cost minimization constrained on the latency at the user equipment (UE). The optimal solution is given by a greedy fractional knapsack algorithm. We illustrate that it is sufficient for the MNO to use a small fraction of the SBS to serve the UE while satisfying the latency constraint. For the problem of overall data rate maximization, we numerically show that the power allocation has significant effect on the social welfare of the system.

\end{abstract}


\section{Introduction} \label{section:introduction}
With the rapid development and innovations of mobile networking technologies, the fifth generation (5G) of mobile communications systems is coming and is going to be rolled out around 2020. The traffic volume will be increased to about $50$ billion devices connected to the network \cite{Cisco2017,FP7}. Hence, the 5G system deployment needs to be cost efficient, reliable, and flexible. These are very challenging requirements needing modifications on both radio access network (RAN) and core network \cite{HZhang2017,3GPP}. 

Accordingly, from the network operators' points of view, the mobile network operators (MNOs) will need to come up with new resource management/resource allocation techniques to improve the network capacity and reduce the latency at the user equipments (UEs) \cite{Tachporn_TMC2017}. However, since the high volume of traffic densities comes from indoor environment such as factories, hospitals, and sport arenas, traditional macro cellular networks become insufficient when indoor UEs need more specific and fast services. In the current network architecture which is dominated by the MNOs, various services cannot be served. Therefore, the wireless systems have to be modified in order to respond rapidly to each specific type of local traffic requirement, i.e., ultra-reliable low latency (uRLLC) services, augmented reality (AR), massive machine type communications and enhanced mobile broadband (eMBB) \cite{Liang2015,Tachporn_TCOM2018}. One possible way to address the above issues is to deploy the micro-operator (MO) networks to serve the specific local services \cite{M_micro, Tachporn_micro2017}. In \cite{M_micro}, the MO concept with relation between MO and other stakeholders was proposed. Also, the aspect of spectrum regulation for MO was provided. In \cite{Tachporn_micro2017}, the authors considered spectrum sharing for MO networks in which one buyer MO bought multiple subbands from the regulator. Then, other MOs would rent the subbands from that buyer MO. In \cite{Tachporn_TWC2016}, the many-to-many matching game with externalities was used to model the spectrum sharing between MNOs. The concept of MO networks, network slicing with mobile edge computing (MEC) and latency constraint at the UE were not considered here.

Different from earlier works where only spectrum regulations and spectrum sharing were proposed, we consider the scenario that the MO allocates the slices of wireless resources as a service including licensed subbands to multiple MNOs. Here, the MO installs the small cell base stations (SBSs) and deploys MEC at each SBS. We study the problem of wireless resource slicing allocation where each MNO obtains multiple slices from the MO while each slice is allocated to at most one MNO. The many-to-one matching game theoretical framework is used to formulate the optimization problem so as to maximize the social welfare of the network defined as sum rate of the MNOs. This becomes a combinatorial optimization problem and the Markov Chain Monte Carlo (MCMC) method is used to compute the global solution of social welfare maximization. However, the transmit power of SBS is considered to be a random variable. Also, the SBS is naturally interested in maximizing its long-term expected data rate by optimizing its power strategy. Therefore, we use the Q-learning method to find the optimal power transmission scheme. Furthermore, given the price of infrastructure of the MNO, we consider how much portion of infrastructure (SBS) should be allocated to the UE so as to satisfy the latency constraint. The solution in terms of the fraction of SBS to serve UE is obtained by solving a fractional knapsack problem.
%


\section{System Model} \label{section:systemmodel}
\begin{figure}[h]
\centering
\includegraphics[height=2.8 in, width=2.8 in, keepaspectratio = true]{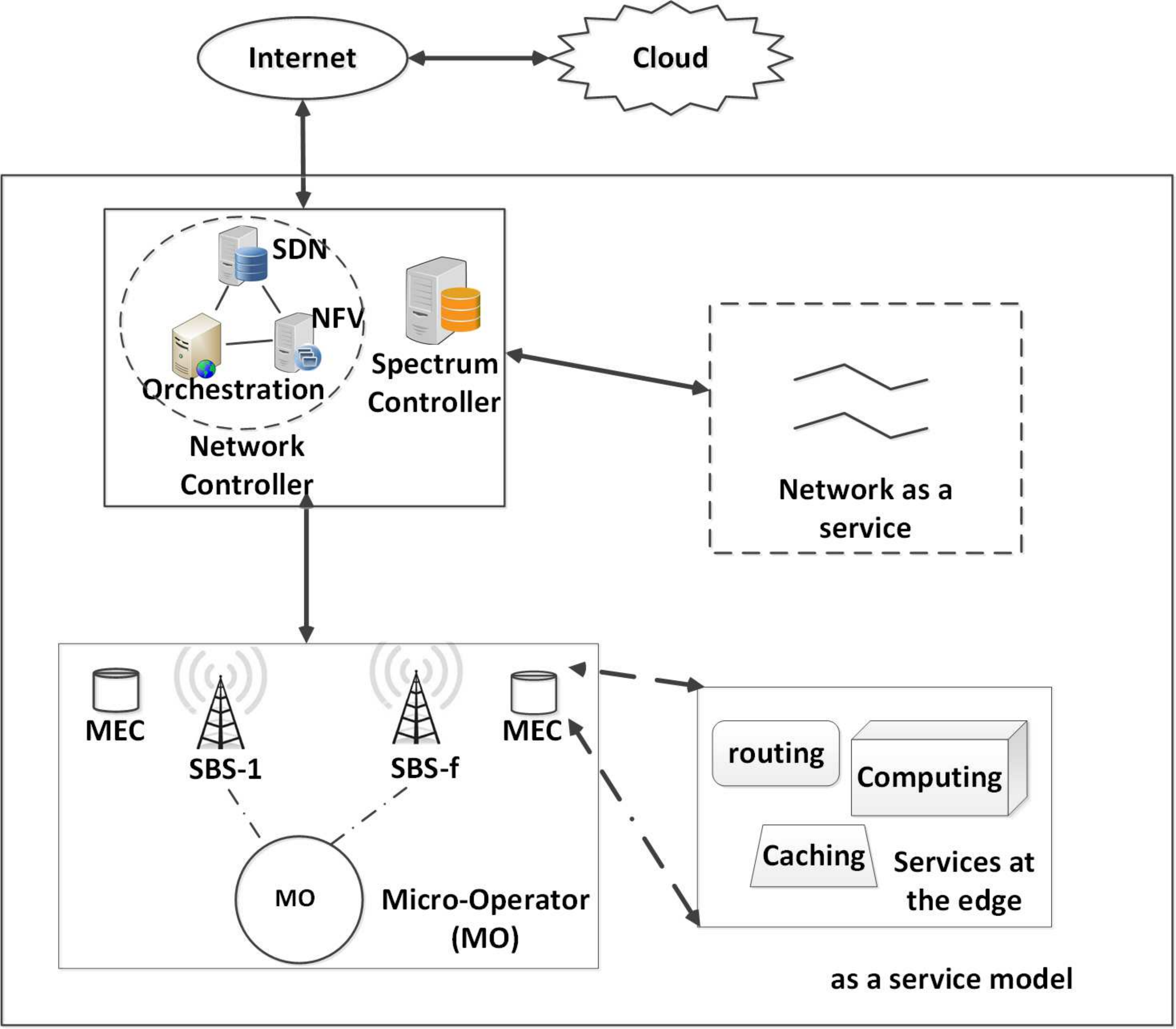}
\caption{System model of network slicing as a service based SDN and NFV for micro-operator (MO) in 5G and B5G}
\label{fig:system_model}
\end{figure}

We consider that the MO installs SBSs and deploys both software-defined networking (SDN) and network function virtualization (NFV). In Figure \ref{fig:system_model}, the MO creates the slices of wireless resource, which consist of network as a service. However, since the MO obtains licensed spectrum from the spectrum controller, the MO will then attach one resource block (RB) in each slice and will allocate to multiple MNOs. At the MEC, multiple functions are placed, i.e., caching in order to cache the popular contents and edge computing server.

Consider the set of $\Kcal$ MNOs with $K$ MNOs serving UEs in the same geographical area. At the initial stage, each MNO-$k\in \Kcal$ obtains SBSs from the MO. Therefore, we focus on slice allocation in which the MO becomes a central controller that match the slices to each MNO. We propose many-to-one matching game framework to model such a scenario where the MNO-$k\in \Kcal$ obtains multiple slices from the MO while each slice can be allocated to at most one MNO. The objective is to maximize the utility of each MNO in terms of the overall achievable rate. Let the set of SBSs subscribed to an MNO-$k$ be given by $\Fcal_k$ with $F_k$ SBSs. Therefore, let $\Fcal = \cup_{k\in\Kcal} \Fcal_k$ be the set of all SBSs in the system. 
The SBS is assumed to employ time division multiple access (TDMA) scheme. Hence each SBS can serve a single UE at a given time slot. Each of the SBSs is assumed to be equipped with a single antenna.

The MNO-$k$ can select any slices in which each slice contains one resource block $l$ (RB-$l$) in $\Lcal_k$ to serve its UEs. Hence, the SBS's transmit power is restricted to a single RB. Let the total power of each SBS be given by $p_{tot}$, which is discretized into $N = \frac{p_{tot}}{\delta}$ levels, where $\delta$ is a quanta of power. Thus, the set of transmit power levels that an SBS-$f$ can choose from is $\Pcal_f = \{0, \delta, 2 \delta, \ldots, (N-1) \delta \}$. We shall denote the transmit power of the SBS-$f$ by $p_f \in \Pcal_f$. The SBSs are assumed to use a probabilistic scheme to select a suitable power level $n \in \{0, 1,\ldots,N-1\}$. Thus, any given action taken by an SBS can be simply represented by $n$.

We further assume that any slice allocated to an MNO can be accessed by more than one UEs. Thus, the data rate of the UE-$f$ associated with SBS-$f$ is given by
\begin{equation}
R_f =  \log_2\Big( 1 + \frac{h_{ff}^{(l)} r_{ff}^{-\alpha} p_{f}}{\sum_{f'\in\mathcal{I}_l} h_{f'f}^{(l)} r_{f'f}^{-\alpha} p_{f'} + \sigma^2 }\Big),
\label{eqn:rate-AP}
\end{equation}
where $p_{f}$ is the transmit power of SBS-$f$ on RB-$l$, $h_{f'f}^{(l)}$ is the channel fading gain between UE-$f$ and SBS-$f'$ using RB-$l$. For simplicity, we assume the fading to be Rayleigh. Also, $\alpha$ denotes path loss exponent and $r_{f'f}$ is the distance between the UE-$f$ and SBS-$f'$. The $\mathcal{I}_l \subset \Fcal$ is the set of SBSs using the same RB-$l$, while $\sigma^2$ is the noise variance. The interference experienced by a UE of an SBS can be categorized as intra-MNO interference. The intra-MNO interference is caused by the fact that the SBSs associated with a given MNO can access any RB attached in each slice assigned to that MNO.

The data rate of MNO-$k$ will be the sum of data rates of each SBS. We can express the rate of MNO-$k$ as,
\begin{equation}
R_{MNO_k}(\Fcal_k, \Lcal_k) = \sum_{f\in\Fcal_k} R_f.
\label{eqn:rate-OP}
\end{equation}

Since the SBS-$f$ can access any one of $L_k$ slices/RBs assigned to its associated MNO-$k$ with equal probability $1/ L_k$, we can express $R_f$ in (\ref{eqn:rate-AP}) as
\begin{equation}
R_f = \frac{1}{L_k}\sum_{l \in \Lcal_k}R_n^{(l)}.
\end{equation}

Let the rate of a downlink SBS-UE system transmitting in a fixed RB-$l$ and at fixed power level $n$ be given by,
\begin{equation}
R_n^{(l)} = \log(1 + SINR_n^{(l)}).
\label{eqn:rate-AP-l-n}
\end{equation}

Here the $SINR_n^{(l)}$ is given by,
\begin{equation}
SINR_n^{(l)} = \frac{h_{ff}^{(l)} r_{ff}^{-\alpha} p_{f}}{\sum_{f'\in\mathcal{I}_l} h_{f'f}^{(l)} r_{f'f}^{-\alpha} p_{f'} + \sigma^2 }.
\label{eqn:SINR-AP-l-n}
\end{equation}

\section{Sum Rate Maximization for Network Slicing} \label{section:BothBSsshareBW}
Consider the social welfare of the network as the overall sum rate as follows:
\begin{equation}
S(\mu) = \sum_{l\in\mathcal{L}} \sum_{k\in\Kcal} x_{lk}  R_{OP_k}(\Fcal_k, \Lcal_k),
\label{eq:socialwelfare1}
\end{equation}
where $\tbf{X} = |\mathcal{L}|\times|\Kcal|$ is a matching matrix $\{x_{lk} : (l,k) \in \mathcal{L} \times \Kcal \}$. We denote the matrix $\tbf{X}$ as,
\begin{equation}
x_{lk} = \left\{
\begin{array}{cl}
1 & \mbox{if}\; \mu(\mbox{MNO}_k) = \mbox{RB}_l \\
0 & \mbox{otherwise}
\end{array}
\right.
\end{equation}
where $\mu$ is a matching.

The objective of the matching game for allocating the slice to multiple MNOs is to maximize the overall data rate. Thus, the optimization problem can be expressed as,
\begin{align}
& S^*(\mu) = \max_{\tbf{X}} \sum_{l\in\mathcal{L}} \sum_{k\in\Kcal} x_{lk} R_{MNO_k}(\Fcal_k, \Lcal_k), \nonumber\\
\mbox{s.t.} \quad\quad &  \mbox{(C1)} \quad \sum_{l\in\Lcal} x_{lk} \leq 1 \quad \forall l \in \mathcal{L}, \label{eqn:centralproblem} \\
& \mbox{(C2)} \quad \sum_{k\in\Kcal} x_{lk} \leq c_k \quad \forall k \in \Kcal.
\end{align}

Constraint $\mbox{(C1)}$ assures that each slice can be allocated to at most one MNO, and constraint $\mbox{(C2)}$ guarantees that each MNO-$k$ can receive at most $c_k$ slices. The MNOs are assumed to be able to communicate with each other through an MO during the slice allocation process, the matching game is used to model the decision process of the MNOs.

\tbf{Algorithm~\ref{alg:mcmc}} proceeds to optimize the social welfare $S$ via the Markov Chain Monte Carlo (MCMC) method. We first initialize with a random matching, and at each iteration, we proceed to accept a swap of random pair of MNOs based on the probability that depends on the change in social welfare. The algorithm keeps track of the best matching found thus far. This algorithm converges to a two-sided exchange-stable matching \cite{ARoth_1992}.


\begin{algorithm}
\caption{MCMC Swap Algorithm}
\label{alg:mcmc}
 \begin{algorithmic}[1]
  \STATE Initialize the matching matrix $\mathbf{X}$.
  \STATE Compute the initial data rate of each MNO-$k$.
 	\FORALL{$t \leq $ maxIterations}
 		\STATE Select a random pair of RBs $\{l, l'\}$.
        \STATE Swap the two RBs for each MNO $\{k, k' \}$ to obtain $\mu_{k}^{k'}$.
        \STATE Update the data rate of MNOs \{$k, k'$\}, $k,k' \in K$, with the Q-learning in \textbf{Algorithm~\ref{alg:Qlearning}}.
        \STATE Compute the social welfare $S_{t}(\mu)$ in (\ref{eq:socialwelfare1}).
        \STATE Compute the transition probability $P_{T_b} = \frac{1}{1 + e^{-T_{b}(S(\mu_k^{k'}) - S(\mu))}}$.
            \IF {$rand() < P_{T_b}$}
            \STATE  $\mu \leftarrow \mu_k^{k'}$ and $S_{t}(\mu) \leftarrow S_{t}(\mu_k^{k'})$
            \ELSIF{$S_{t}(\mu) > S_{t-1}(\mu_{k}^{k'})$}
 		    \STATE $\mu \leftarrow \mu_k^{k'}$
            \STATE Update the social welfare $S_{t}(\mu) \leftarrow S_{t}(\mu_k^{k'})$.
 	        \ENDIF
        \STATE $t \leftarrow t+1$.
    \ENDFOR
 \end{algorithmic}
 \end{algorithm}

\subsection{Delay from the Computation Files at MEC for one MNO}
\label{subsec:MEC_schedule}
Since each SBS is assumed to deploy MEC, we also consider the computation of the files at each SBS. Let the requested files arriving at the MEC server be a Poisson process with exponentially distributed inter-arrival times, and the arrival rate $\lambda$. At the MEC server, the time is partitioned into multiple time slots, with the length $Q_m$ seconds per slot, and the files are scheduled using round-robin method. Let the computing capability of the MEC server be $s_m$ in cycles/second and $\tau$ be CPU cycles/bit. Then, in order to finish processing file $x_f$ of the UE-$f$ associated with SBS-$f \in \Fcal_k$, the number of time slots needed is $n = \ceil[\bigg]{\frac{\tau x_f}{s_m Q_m}}$. Hence, the required service time $D_{s,m}$ to complete processing the file in seconds is,
\beq
D_{s,m}  = n Q_m.
\label{eq:MEC_delay_queue}
\eeq

Each UE has file size $x_f$ bits to be computed at the MEC, and the file $x_f$ bits use $T_s$ seconds to finish the computation.  Therefore, the SBS-$f \in \Fcal_k$ will transmit signal and the computed file to UE-$f$ and the delay can be considered in terms of downlink channel capacity as,
\beq
D_{ff}^{d} = \frac{x_f}{R_n^{(l)}T_s}
\label{eq:schedule_delay}
\eeq
where $R_n^{(l)}$ is obtained by computing $Q$-learning in \tbf{Algorithm~\ref{alg:Qlearning}}. The total delay includes the service time delay from processing the file $D_{s,m}$ and the delay from downlink transmissions $D^d_{ff}$, which can be expressed as,
\beq
D_f = D_{s,m} + D_{ff}^d
\label{Eqn:DelayD_f}
\eeq
where $D_{s,m}$ and $D_{ff}^d$ are given by (\ref{eq:MEC_delay_queue}), and (\ref{eq:schedule_delay}), respectively.
%

\subsection{Infrastructure Cost Minimization with Latency Constraint at the UE}
\label{subsec:MEC_schedule}
In this part, we consider one MNO and formulate the cost of infrastructure minimization problem with the latency constraint at the UE. The problem can be written as a linear program as follows:
\begin{align}
\label{eqn:knapsack}
& \min_{y_{fk}} \sum_{f\in \Fcal_k} c_f y_{fk}, \\
\mbox{s.t.} &  \quad \mbox{(C1)} \quad y_{fk} \Pr(D_f\geq D_{th})
\leq \epsilon.
\label{eqn:Prob_Delay}
\end{align}
The constraint $\mbox{(C1)}$ in (\ref{eqn:Prob_Delay}) is a probabilistic delay constraint that ensures the latency is bounded by a threshold value $D_{th}$ with a probability $\epsilon \in (0,1)$. The $c_k$ is the price of infrastructure (SBSs) when each SBS is utilized and $y_{fk} (0 \leq y_{fk} \leq 1)$ denotes the fraction of infrastructure when the UE-$f$ is served. To make the problem more tractable, we have from Markov's inequality
\beq
\text{Pr}(D_f  \geq  D_{\text{th}}) \leq \frac{\Ebb[D_f]}{ D_{\text{th}}} \leq \epsilon.
\label{eqn:Markov_in_Equal}
\eeq

Using the Markov's inequality (\ref{eqn:Markov_in_Equal}), we can linearize the probabilistic constraint in (\ref{eqn:Prob_Delay}) as
$\Ebb[D_f] \leq \epsilon D_{\text{th}}$.
Since we can express
\beq
\Ebb[D_f] = \Ebb[D_{s,m} + D_{ff}^d] = D_{s,m} + D_{ff}.
\label{Expected_Delay}
\eeq
Substituting (\ref{Expected_Delay}) in (\ref{eqn:Markov_in_Equal}) We can rewrite the constraint (C1) as
\beq
(C1') \quad  y_{fk}( D_{s,m} + D_{ff}^d ) \leq \epsilon D_{\text{th}}.
\eeq

The problem (\ref{eqn:knapsack}) is an instance of knapsack problem. The SBSs are interpreted as ``items'', the delay is interpreted as ``weights'', and the right hand term of constrain (C1) in (\ref{eqn:knapsack}) is interpreted as ``weight capacity'' of a bag. Since $y_{fk} \in [0,1]$, the problem (\ref{eqn:knapsack}) becomes a \tit{fractional knapsack problem} and a greedy algorithm can be used to obtain the optimal solution \cite[Chap 17.1]{Korte2012}. The greedy algorithm is provided in \tbf{Algorithm~\ref{alg:FracKnapsack}}. The idea behind this greedy algorithm is as follows. We first sort the SBSs according to the cost of each SBS in an ascending order. We then assign $y_{fk} = 1$ if the weight (the total delay $D_f$ in (\ref{Eqn:DelayD_f})) is less than or equal to the residual weight capacity of knapsack. In our case, the maximum weight capacity of knapsack is defined by $\bar{w} = \epsilon D_{th}$

\beq
y_{fk} = \left\{\begin{array}{cr}
				1, & \mbox{if} \quad D_f \leq \bar{w}-w  \\
				\frac{(\bar{w}-w)}{D_f}, & \mbox{if} \quad D_f > \bar{w}-w,
			 \end{array} \right.
\eeq
where $w$ is the weight in the knapsack thus far.
\begin{algorithm}
\caption{Fractional Knapsack Algorithm}
\label{alg:FracKnapsack}
 \begin{algorithmic}[1]
 \STATE Initialize $y_{fk} =0$, $w = 0$, and $V = 0$.
 \STATE Compute $R_n^{(l)}$ by using $Q$-learning in \tbf{Algorithm~\ref{alg:Qlearning}} and then substitute the obtained $R_n^{(l)}$ in (\ref{eq:schedule_delay}).
 \STATE Calculate $D_f$ using (\ref{Eqn:DelayD_f}).
 \STATE Compute $\rho_f= c_f/D_f$.
  \STATE Sort $\rho_f$ in ascending order such that $\rho_{\pi_1} \leq \rho_{\pi_2} \leq \cdots \leq \rho_{\pi_{F_k}}$.
    \FOR{$i = 1$ \TO $F_k$}
        \IF{$D_{\pi_i} \leq \bar{w} - w$}
            \STATE $y_{\pi_i} = 1$
             \STATE $V = V + c_{\pi_i}$
		\STATE $w = w + D_{\pi_i}$
         \ELSE
            \STATE $y_{\pi_i} = \frac{\bar{w}-w}{D_{\pi_i}}$
            \STATE $V = V + c_{\pi_i} y_{\pi_i}$
		\STATE Terminate
      \ENDIF
 \ENDFOR
 \end{algorithmic}
 \end{algorithm}

\section{Self-organizing SBSs using Reinforcement Learning Strategy}\label{section:Reinforcementlearning}
In this section, we propose a mechanism of self-organizing networks based on reinforcement learning. We assume that all the SBSs are able to estimate the interference they experience at each RB and accordingly tune their transmission strategies towards a better performance based on Q-learning.

\subsection{$Q$-learning}
The $Q$-learning model consists of a set of states $\Scal$ and actions $\Acal$ aiming at finding a policy that maximizes the observed rewards over the interaction time of the agents/players (i.e., small cells). Every slice with SBS $f \in \Fcal_k$ allocated to an MNO-$k$, where $k \in \Kcal$  explores its environment, observes its current state $s$, and takes a subsequent action $a$, according to a decision policy $\pi: s \rightarrow a$.

For each MNO-$k$, let us denote by $\mathcal{G}_k^{Q}=\big(\Fcal_k,\lbrace \mathcal{P}_f \rbrace_{f\in \Fcal_k},\lbrace u_f \rbrace_{f\in \Fcal_k}\big)$ the $Q$-learning game. Here, the players of the game are the SBSs $f \in \Fcal_k$ which seek to allocate power in the RBs assigned to their corresponding MNO. The $s_f(t)$ is the state of SBS-$f$ at time $t$. The state of an SBS is a binary variable, $s_f(t) \in \{0,1\}$, which indicates whether SBS-$f$ experiences interference in RB-$l$ assigned to its corresponding MNO-$k$ such that its required QoS is violated. The QoS requirement is said to be violated when $SINR_n^{(l)} < SINR_{th}$, where $SINR_n^{(l)}$ is given by (\ref{eqn:SINR-AP-l-n}). The $a_f(t)$ is the action of SBS-$f$, where $a_f(t) \in \Pcal_f$. Any given action can be represented by an integer variable $a_f (t) \equiv n$, where $n$ represents the power level. Finally, $u_f(t)$ is the utility function or payoff of SBS-$f$ at time-instant $t$, which we take as the instantaneous rate of SBS-$f$ at time-instant $t$ as given by (\ref{eqn:rate-AP-l-n}) if the QoS is satisfied. Otherwise it is taken to be zero:
\begin{equation}
u_f(t) = \left\{
\begin{array}{cl}
R_n^{(l)} & \mbox{iff}\; SINR_n^{(l)} \geq SINR_{th} \\
0 & \mbox{otherwise}.
\end{array}
\right.
\end{equation}

The \emph{expected} discounted reward over an infinite horizon can be given by:
\begin{equation}
 V^{\pi}(s)=W(s,\pi^*(s)) +\gamma \sum_{v \in S} P_{s,v}(\pi(s))V^{\pi}(v),
\label{eq:VI}
 \end{equation}
where $0\leq \gamma \leq 1$ is a discount factor and $r$ is the agent's reward at time $t$. $W(s,\pi^*(s))=\Ebb\{w(s,\pi(s))\}$ is the mean value of reward $w(s,\pi(s))$, and $P_{s,v}$ is the transition probability from state $s$ to $v$. For a given policy $\pi$, we can define a $Q$-value as:
 \begin{equation}
 Q^*(s,a)=W(s,a)+\gamma \sum_{v \in S} P_{s,v}(a)V^{\pi}(v),
 \label{eq:VIII}
 \end{equation}
 which is the expected discounted reward when executing action $a$ at state $s$ and then following policy $\pi$ thereafter. The actions are chosen according to their $Q$-values as:
\begin{equation}
 P(a|s)= \frac{e^{Q(s^k,a)/T_p}}{\sum_{a' \neq a}e^{Q(s^k,a')/T_p}}.
 \end{equation}

The $Q$-learning process aims at finding $Q(s,a)$ in a recursive manner where the update equation is given as \cite{Fudenberg1998}:
\begin{align}
Q_{t+1}(s_t,a_t) &= (1-\beta_t)Q_{t}(s_t,a_t) + \nonumber\\
                         &\beta_t \left[w(s_t,a_t)+ \gamma \max_{a_{t}'\neq a_t}Q_{t}(s_t,a_{t}')\right],
\label{eqn:Qupdate}
\end{align}
where $\beta_t$  is the learning rate such that $0\leq\beta_t<1$. The Q-learning algorithm for power allocation at each SBS-$f$ is described in \tbf{Algorithm~\ref{alg:Qlearning}}.

\begin{algorithm}
\caption{$Q$-learning algorithm for power allocation}
\label{alg:Qlearning}
\begin{algorithmic}[1]
 \STATE $Q(s,a)=0$
        \FORALL {$t \leq $ maxIterations}
        \FOR {$k = 1:K_{aug}$}
        \STATE Calculate the utility $u_f$.
            \IF {$rand() \leq \gamma$}
            \STATE Randomly choose an action (power level) $n$.
            \ELSE
            \STATE Choose a state with $n^{*} = \text{argmax}_{n}Q(s,a)$.
            \ENDIF
        \STATE Each SBS-$f$ computes the expected date rate ($R_f$).
        \STATE Update $Q$-value $Q_{t+1}(s_t,a_t)=(1-\beta_t)Q_{t}(s_t,a_t)+ \beta_t \left[w(s_t,a_t)+ \gamma \max_{a_{t}'\neq a_t}Q_{t}(s_t,a_{t}')\right]$.
        \STATE $t \leftarrow t+1$
        \ENDFOR
        \ENDFOR
 \end{algorithmic}
 \end{algorithm}

\section{Simulation Results} \label{section:Results}
In this section, we present numerical results to evaluate the performance of network slicing allocation, the cost of infrastructure minimization and proposed algorithms. Each MNO is assumed to have $8$ SBSs per $\pi \times 500^2$ square meters. We consider TDMA system and thus, the SBS serves a single UE in a particular time slot and each UE is located within 20 meters of the SBS. The direct path loss between SBS and SBS-UE at distance $d$ meters is given by $\text{PL}(d) = 37 + 20\text{log}_{10}(d)$ dB, and the path loss due to wall, $\text{PL}_{\text{wall}} = 15$ dB. The standard deviation of log-normal shadow fading is assumed to be $4$ dB. The cross gain path loss between SBS and SBS-UE at distance $d_{S-UE}$ is given by $\text{PL}(d_{S-UE}) = 7+56\text{log}_{10}(d_{S-UE}) + \text{PL}_{\text{wall}}$. The maximum transmit power of each SBS is $10$ dBm, and the noise variance is $-120$ dBm. The SINR threshold at each UE is $3$ dB. The temperature $T_b$ in MCMC swap algorithm is $100$. In the cost of infrastructure minimization problem, we assume that the size of the file of UE-$f$ is $x_f = 100$ bits, the computing capability of MEC server is $s_m = 20$ cycles/second, $\tau = 15$ CPU cycles/bit and the time in MEC server is $Q_m = 0.9$ secs/slot. The price of $8$ SBSs is $\mathbf{c}_f = [50, 80, 200, 500, 800, 1000, 300, 400]$. In the Q-learning algorithm, we set the parameters as follows: discount factor $\gamma = 0.95$, and the learning rate $\beta_t = 0.5$. We run 2500 iterations for MCMC swap algorithm and $2000$ instances for Q-learning algorithm.

\begin{figure}[h]
\centering
\includegraphics[height=3.2 in, width=3.3 in, keepaspectratio = true]{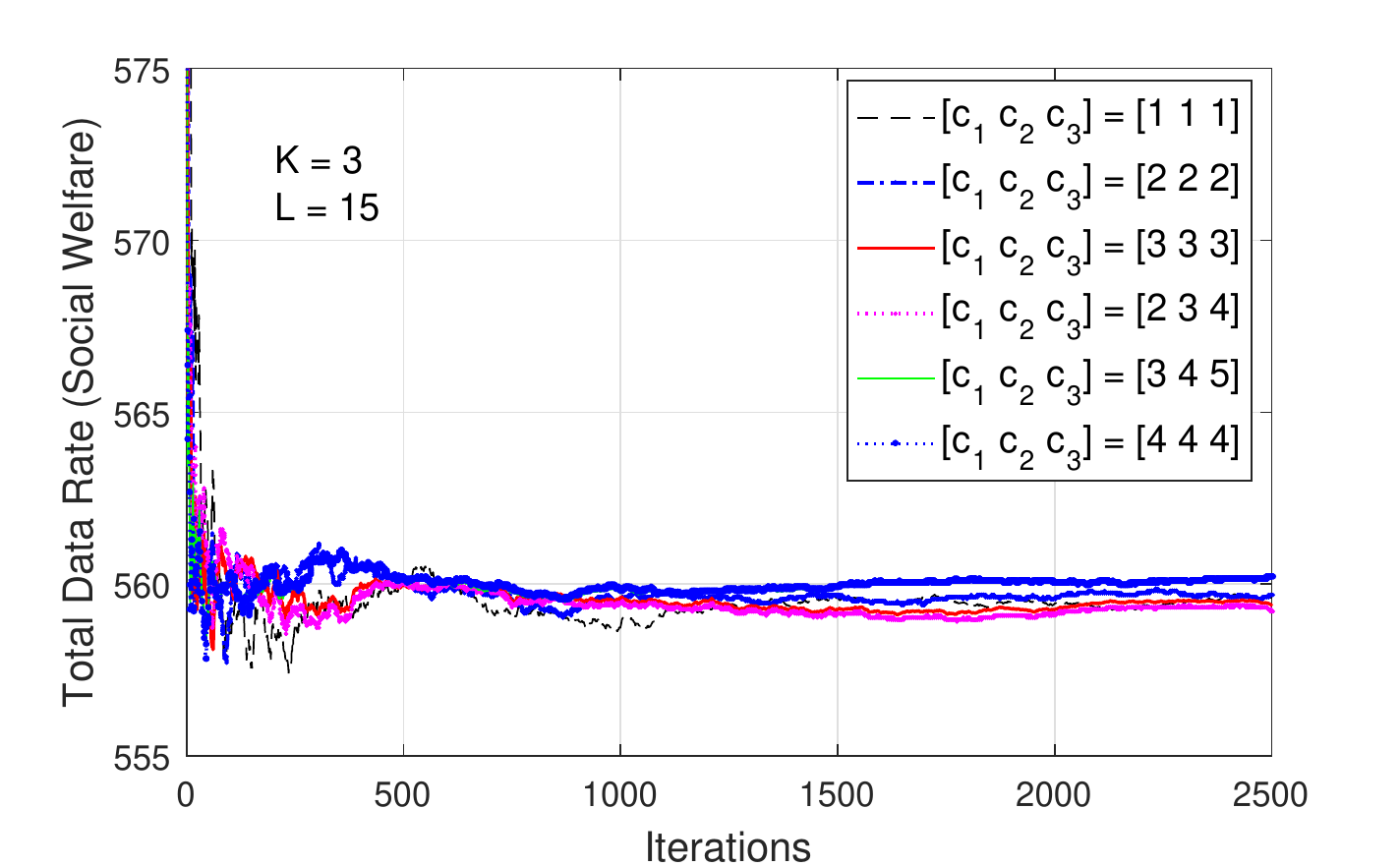}
\caption{Convergence of social welfare for $K = 3$ MNOs using MCMC swap with Q-learning power allocation}
\label{fig:Change_c}
\end{figure}

In Fig.~\ref{fig:Change_c}, we show the convergence of the social welfare (bits/sec/Hz) using MCMC swap algorithm when there are $K = 3$ MNOs, the number of slices/RBs is $L = 15$ while using $Q$-learning for power allocation. We see that the system converges to the steady state. At the steady state, we can observe that changing the number of slices/RBs $\mathbf{c}_k$ obtained by each MNO-$k$ does not have much effect on the social welfare.

\begin{figure}[h]
\centering
\includegraphics[height=3.2 in, width=3.3 in, keepaspectratio = true]{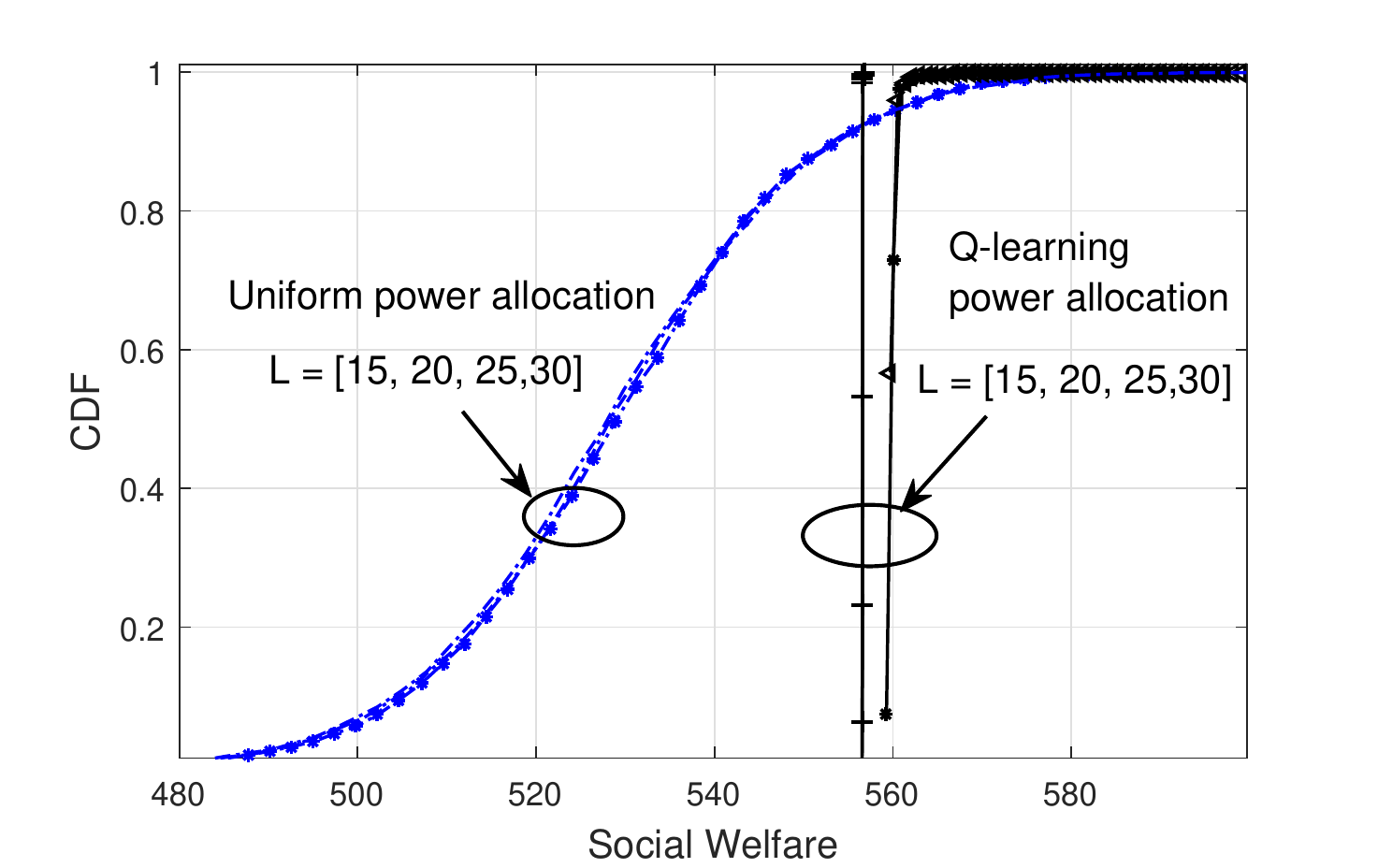}
\caption{Comparison of the cumulative distribution function (CDF) of social welfare when changing number of slices/RBs ($L$) while using $Q$-learning and uniform power allocation schemes}
\label{fig:Uniform_Qlearning}
\end{figure}

In Fig.~\ref{fig:Uniform_Qlearning}, the cumulative distribution function (CDF) of the overall social welfare for $K = 3$ MNOs, the number of slices/RBs is assumed to be $L = [15, 20, 25, 30]$ and the maximum number of slices allocated to each MNO-$k$, $k = \{1,2,3\}$ is $[c_1, c_2, c_3] = [2,3,4]$. We consider cases when each SBS allocates power to its UE using Q-learning and uniform power allocation. We see that different power allocation scheme significantly affect the system performance compared with changing the number of slices/RBs. The Q-learning power allocation gives much higher social welfare than that of the uniform power allocation. Therefore, for a given number of MNOs, the effect of power allocation is much more significant than the effect of slice allocation for the social welfare of the system.

\begin{figure}[h]
\centering
\includegraphics[height=3.2 in, width=3.3 in, keepaspectratio = true]{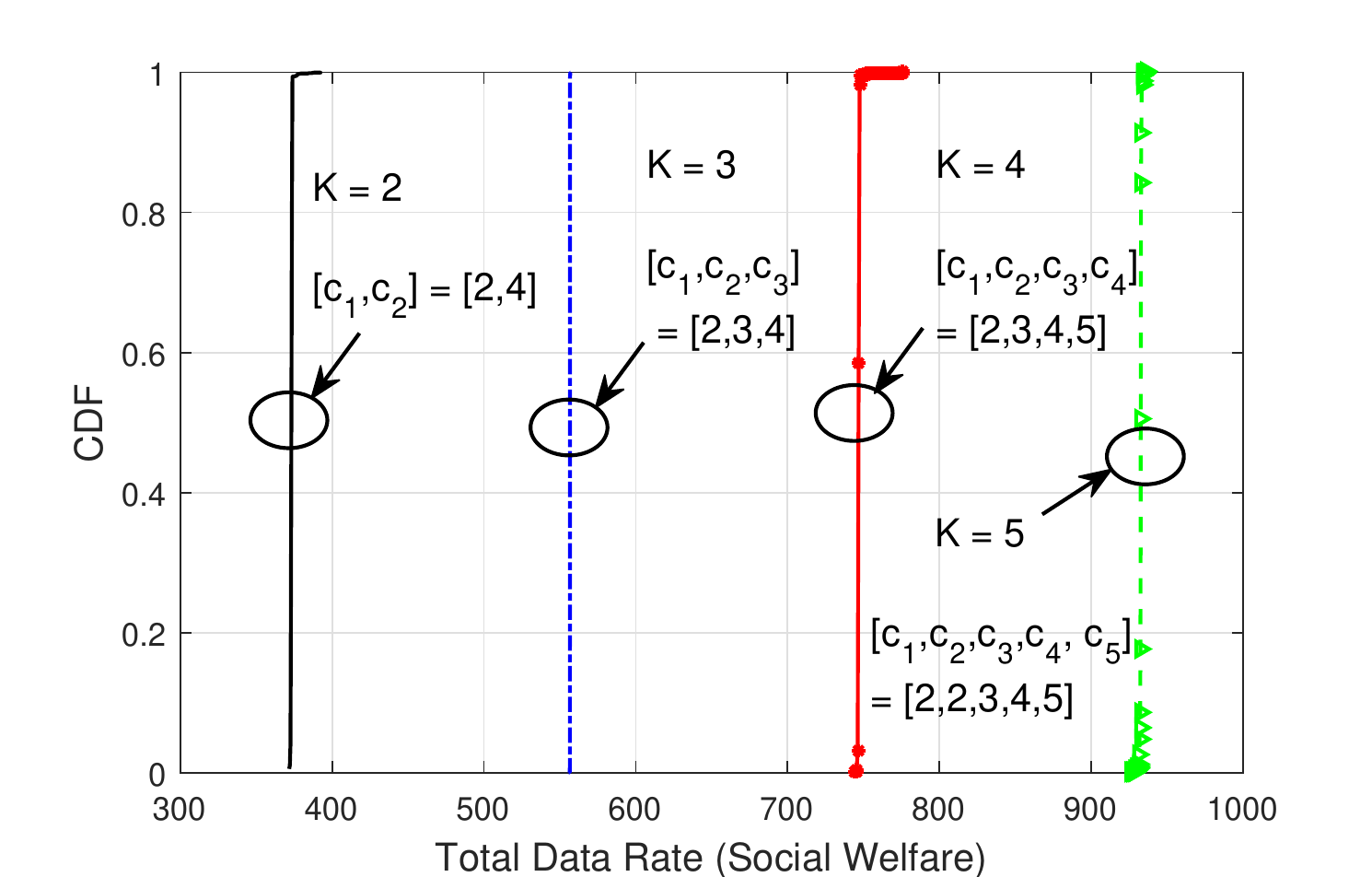}
\caption{Comparison of the cumulative distribution function (CDF) of social welfare for $K = 3,4,5$ MNOs}
\label{fig:CDF_MNOs}
\end{figure}

In Fig.~\ref{fig:CDF_MNOs}, we set $L=15$ and plot the CDF of the overall social welfare while varying the number of MNOs. Each SBS uses the Q-learning scheme for power allocation. We can see that when increasing the number of MNOs, the social welfare is enhanced significantly.

\begin{figure}[h]
\centering
\includegraphics[height=3.5 in, width=3.5 in, keepaspectratio = true]{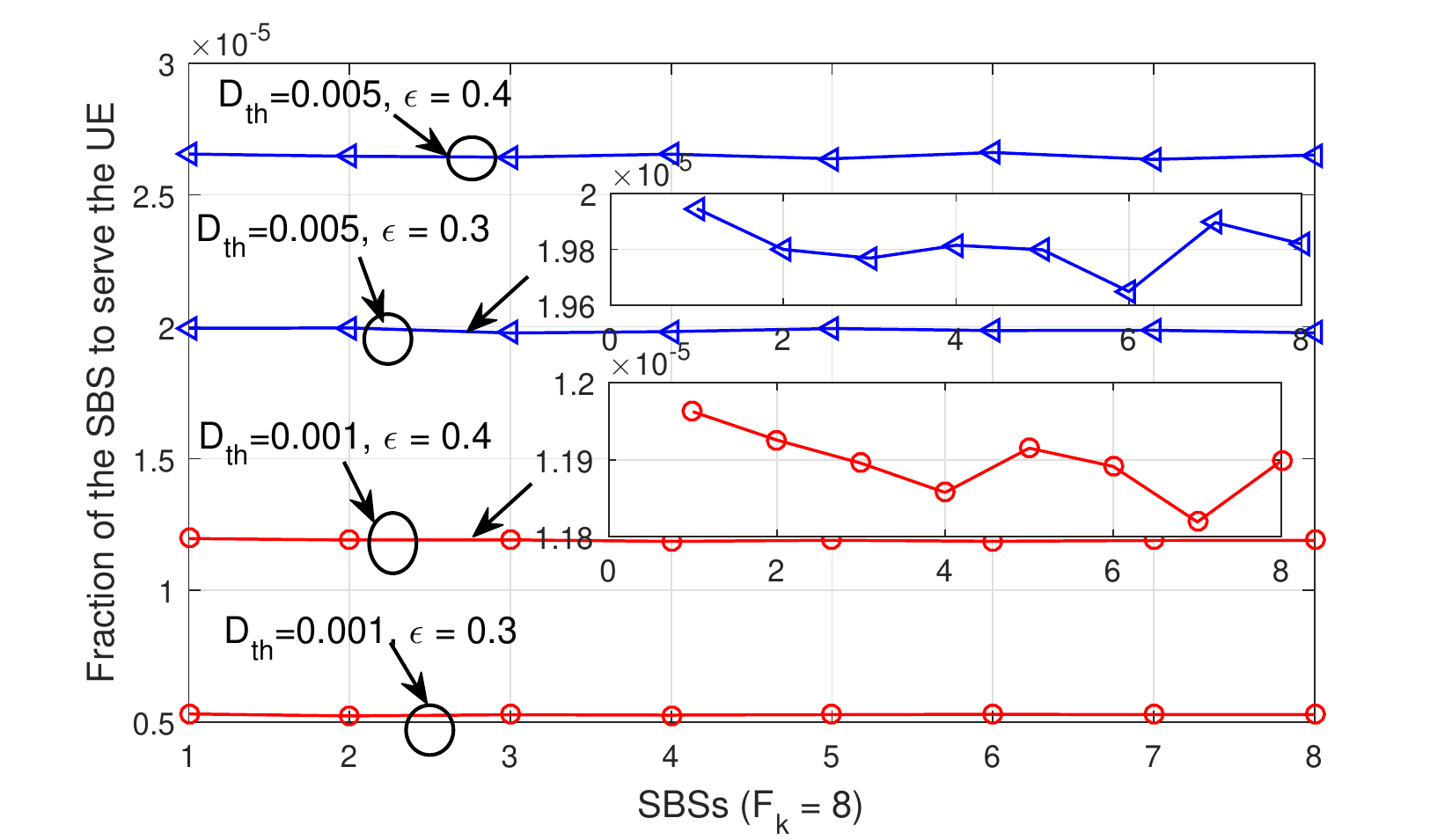}
\caption{The fraction of SBSs (variable $y_{fk}$) versus SBSs while changing the delay threshold $D_{\text{th}}$ and the tolerable probability $\epsilon$}
\label{fig:fraction_SBSs}
\end{figure}

\begin{figure}[h]
\centering
\includegraphics[height=3.4 in, width=3.4 in, keepaspectratio = true]{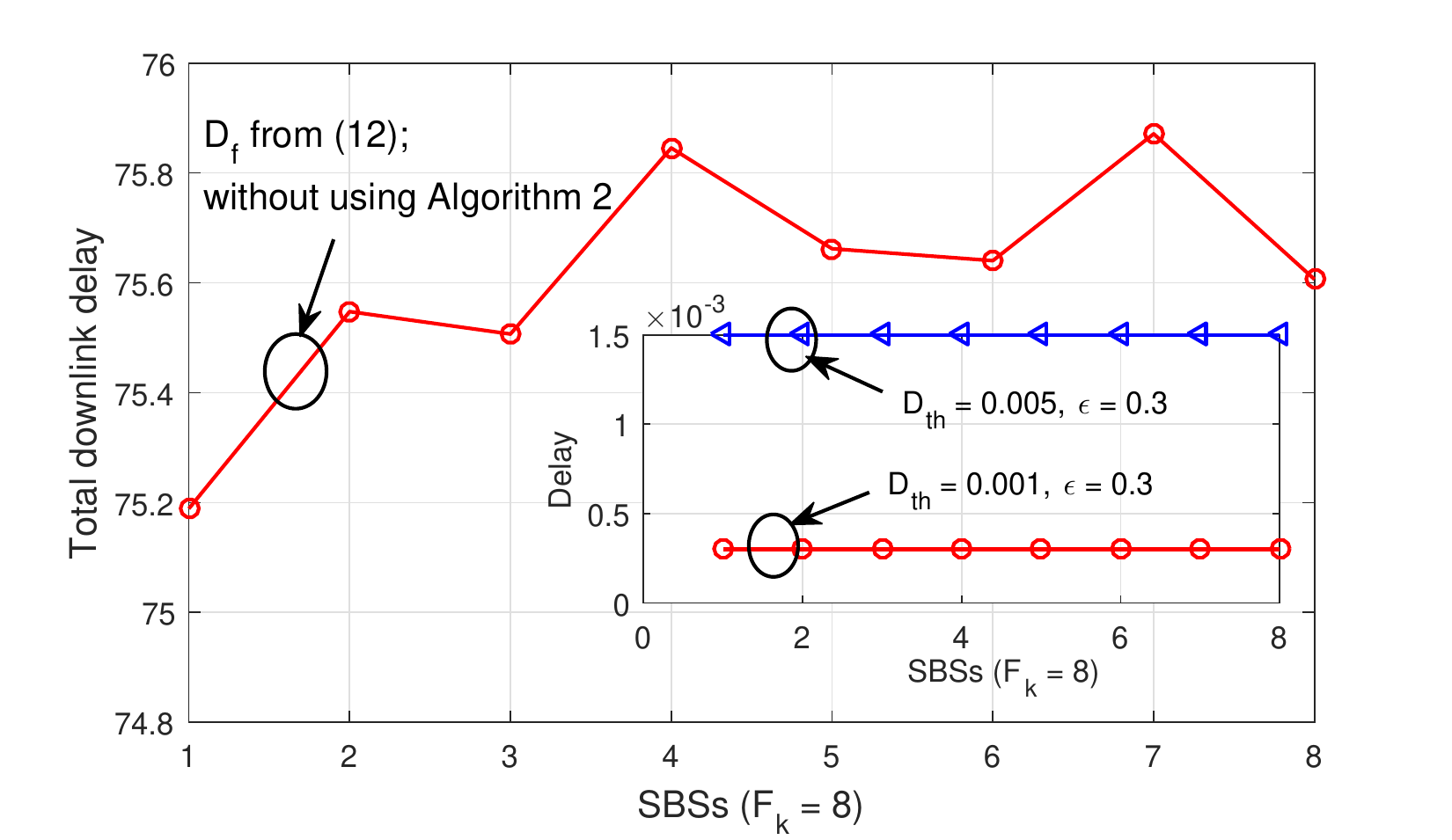}
\caption{The total delay from downlink transmission and computing the files}
\label{fig:tot_delay}
\end{figure}

In Fig.~\ref{fig:fraction_SBSs} and Fig.~\ref{fig:tot_delay}, we plot the results from the infrastructure cost minimization problem in (\ref{eqn:knapsack})-(\ref{eqn:Prob_Delay}). Each SBS of the MNO-$k$ determines the transmit power by using the $Q$-learning scheme. We assume that the latency is bounded by a threshold value $D_{\text{th}} = 0.001$ and $0.003$ while changing the value of the tolerable probability as $\epsilon = 0.3, 0.4$. The fraction of infrastructure (variable $y_{fk}$) versus SBSs is illustrated in Fig.~\ref{fig:fraction_SBSs}. The fractional variable $y_{fk}$ indicates the portion of infrastructure (SBS-$f\in\Fcal_k$) that the MNO-$k$ uses to serve the UE. We see that when $\epsilon$ and $D_{th}$ increase, the fraction of SBS which is used to serve the UE is also increased.

In Fig.~\ref{fig:tot_delay}, we show the total transmission delay $D_f$ from (\ref{Eqn:DelayD_f}) versus SBSs without using \tbf{Algorithm 2}. We also illustrate the total delay after using the \tbf{Algorithm 2} in a small figure. The tolerable probability is $\epsilon = 0.3$ and the threshold $D_{\text{th}} = 0.001$ and $0.005$. From Fig.~\ref{fig:fraction_SBSs} and Fig.~\ref{fig:tot_delay}, we can observe that it is sufficient for the MNO-$k$ to use a small fraction of each SBS to serve the UE in order to satisfy the latency constraint at the UE.

\section{Conclusion}
\label{subsec:conclusion}
We have modeled the network slicing allocation with the mobile edge computing deployment for micro-operator (MO) networks. The MO has created the slices of wireless resource and then allocated these slices to multiple mobile network operators (MNOs). We have formulated the optimization problem to maximize the social welfare, defined as sum rate of all MNOs. The many-to-one matching game has been used to obtain the global optimal solution of the problem. The results have been computed by using Markov Chain Monte Carlo algorithm. Also, the Q-learning method has been proposed to obtain the optimal random transmit power strategy of the small cell base stations (SBSs). Furthermore, for an MNO, we have explored the problem of infrastructure cost minimization constrained on the latency at each user equipment (UE). The solution of the minimization of infrastructure cost has been given by a greedy fractional knapsack algorithm. We have observed that the MNO can use a small fraction of SBS to serve UE so as to satisfy the latency constraint at the UE. For the problem of maximization of social welfare, we have shown numerically that the results are stable and socially optimal. One of the important conclusions that we could highlight is that the power allocation has greater effect on the social welfare than that of slice allocation. The proposed framework can be enhanced by considering multiple MOs deployment in factory and hospital. The MOs can be assumed to serve machine type communications in addition to mobile broadband services. This direction would be an interesting extension of this work since very low latency and reliability will need to be considered.

\section*{Acknowledgment}
This work has been financially supported by 6Genesis (6G) Flagship project (grant 318927).

\bibliographystyle{IEEE}

\begin{thebibliography}{1}

\bibitem{Cisco2017}
Cisco, ``Cisco visual networking index : forecast and methodology, 2016-2021,'' {\em White Paper}, Jun., 2017.

\bibitem{FP7}
FP7 European Project 317669 METIS, ``Mobile and Wireless Communications Enablers for the Twenty-Twenty Information Society 2012,'' {\em [Online]. Available: https://www.metis2020.com/}.


\bibitem{HZhang2017}
H.~Zhang, {\em et al.}, ``Network slicing based 5G and future mobile networks: mobility, resource management, and challenges,'' {\em IEEE Commun. Magazine}, pp. 138-145, vol. 55, 2017.


\bibitem{3GPP}
3rd Generation Partnership Project (3GPP), ``Digital Cellular Telecommunications System (Phase 2+)({GSM}); Universal Mobile Telecommunications System ({UMTS}); LTE; Service aspects and requirements for network sharing (3{GPP} {TR} 22.951
                  version 14.0.0 Release 14),'' 2017.


\bibitem{Tachporn_TMC2017}
T.~Sanguanpuak, {\em et al.}, ``Infrastructure sharing for mobile network operators: analysis of trade-offs and market,'' {\em IEEE Trans. on Mobile Computing}, 2018.


\bibitem{Liang2015}
C.~Liang and F.~R.~Yu, ``Wireless network virtualization: a survey, some research issues and challenges,'' {\em IEEE Commun. Surveys Tutorials}, pp. 358-380, vol. 17, 2015.


\bibitem{Tachporn_TCOM2018}
T.~Sanguanpuak, {\em et al.}, ``Edge Caching in Delay-Constrained Virtualized Cellular Networks : Analysis and Market,'' {\em arXiv:1802.04769v1 [cs.IT] 13 Feb 2018}.


\bibitem{M_micro}
M.~Martinmikko, {\em et al.}, ``Micro-operators to boost local service delivery in 5G, {\em Wireless Personal Communications, Springer}, pp.69-82, Jul. 2017.


\bibitem{Tachporn_micro2017}
T.~Sanguanpuak, {\em et al.}, ``On spectrum sharing among micro-operators in 5G,'' {\em IEEE European Conference on Networks and Communications (EuCNC)}, pp. 1-6, 2017.




\bibitem{Tachporn_TWC2016}
T.~Sanguanpuak, {\em et al.}, ``Multi-Operator Spectrum Sharing for Small Cell Networks : A Matching Game Perspective,'' {\em IEEE Trans. on Wireless Communication}, 2016.


\bibitem{ARoth_1992}
A.~Roth, and M.A.O.~Sotomayor, {\em Two-Sided Matching: A Study in Game Theoretic Modeling and Analysis}, Cambridge Press, 1992.


\bibitem{Korte2012}
B.~Korte and J.~Vygen, {\em Combinatorial Optimization: Theory and Algorithms}, 5th ed., Springer, 2012.


\bibitem{Fudenberg1998}
D.~Fudenberg and D.K.~Levine, {\em The Theory of Learning in Games}, Cambridge, MA:MIT Press, 1998.




\end{thebibliography}

\end{document}